\documentclass[sigconf]{acmart}

\usepackage{todonotes}

\usepackage{amsmath}
\DeclareMathOperator*{\argmax}{arg\,max}

\AtBeginDocument{%
  \providecommand\BibTeX{{%
    \normalfont B\kern-0.5em{\scshape i\kern-0.25em b}\kern-0.8em\TeX}}}

\setcopyright{acmcopyright}
\acmDOI{10.1145/1122445.1122456}

\acmConference[GECCO '23]{The Genetic and Evolutionary Computation Conference 2023}{July 15--19, 2023}{Lisbon, Portugal}
\acmYear{2023}
\copyrightyear{2023}

\newtheorem{dfn}{Definition}
\newcommand{\onlysat}{\textup{\textsc{sat}}}
\newcommand{\onlytsat}{\textup{\textsc{3sat }}}
\newcommand{\onlytsatt}{\textup{\textsc{3sat}}}
\newcommand{\onlyand}{\textup{\textsc{AND}}}
\newcommand{\onlyor}{\textup{\textsc{OR}}}
\newcommand{\onlynot}{\textup{\textsc{NOT}}}
\begin{document}

\title{Influence of Different 3SAT-to-QUBO Transformations on the Solution Quality of Quantum Annealing: A Benchmark Study}


\author{Sebastian Zielinski}
\affiliation{%
  \institution{LMU Munich}
  \streetaddress{Oettingenstraße 67}
  \city{Munich}
  \country{Germany}
  }
\email{sebastian.zielinski@ifi.lmu.de}

\author{Jonas Nüßlein}
\affiliation{%
  \institution{LMU Munich}
  \streetaddress{Oettingenstraße 67}
  \city{Munich}
  \country{Germany}
  }
\email{jonas.nuesslein@ifi.lmu.de}

\author{Jonas Stein}
\affiliation{%
  \institution{LMU Munich}
  \streetaddress{Oettingenstraße 67}
  \city{Munich}
  \country{Germany}
  }
\email{jonas.stein@ifi.lmu.de}

\author{Thomas Gabor}
\affiliation{%
  \institution{LMU Munich}
  \streetaddress{Oettingenstraße 67}
  \city{Munich}
  \country{Germany}
  }
\email{thomas.gabor@ifi.lmu.de}

\author{Claudia Linnhoff-Popien}
\affiliation{%
  \institution{LMU Munich}
  \streetaddress{Oettingenstraße 67}
  \city{Munich}
  \country{Germany}
  }
\email{linnhoff@ifi.lmu.de}

\author{Sebastian Feld}
\affiliation{%
  \institution{TU Delft}
  \streetaddress{Mekelweg 4}
  \city{Delft}
  \country{Netherlands}
  }
\email{s.feld@tudelft.nl}

\renewcommand{\shortauthors}{Zielinski et al.}

\begin{abstract}
To solve \onlytsat instances on quantum annealers they need to be transformed to an instance of Quadratic Unconstrained Binary Optimization (QUBO). When there are multiple transformations available, the question arises whether different transformations lead to differences in the obtained solution quality. Thus, in this paper we conduct an empirical benchmark study, in which we compare four structurally different QUBO transformations for the \onlytsat problem with regards to the solution quality on D-Wave's \emph{Advantage\_system4.1}. We show that the choice of QUBO transformation can significantly impact the number of correct solutions the quantum annealer returns. Furthermore, we show that the size of a QUBO instance (i.e., the dimension of the QUBO matrix) is not a sufficient predictor for solution quality, as larger QUBO instances may produce better results than smaller QUBO instances for the same problem. We also empirically show that the number of different quadratic values of a QUBO instance, combined with their range, can significantly impact the solution quality.
\end{abstract}




\maketitle

\section{Introduction}
In the last years, quantum computing has garnered much attention from the research community, as the first quantum computers became available. Well known theoretical results show that quantum algorithms can solve hard problems with (potentially) exponential speedup over their classical counterparts \cite{shor1994algorithms}. Thus, it is not surprising that researchers try to develop and employ new quantum algorithms for hard problems in their respective research areas.\\
In this work, we focus on a special type of quantum computers: \emph{quantum annealers}. 
Quantum Annealing is a heuristic approach to solve combinatorial optimization problems. It may be seen as a variation of the more common simulated annealing (SA) metaheuristic \cite{mcgeoch2014adiabatic}. The idea of incorporating quantum phenomena rather than thermal annealing into a heuristic optimization framework appears to have been independently suggested by several researchers (i.e., \cite{apolloni1990numerical, ray1989sherrington, farhi2001quantum, kadowaki1998quantum}) \cite{mcgeoch2014adiabatic}. Commercially available quantum annealers are currently produced by company D-Wave Systems. These devices employ a version of quantum annealing (QA), which is based on the principles of adiabatic quantum computing (see \cite{farhi2000quantum, farhi2001quantum}), to solve Ising spin glass problems \cite{kadowaki1998quantum} of the following kind:
\begin{equation}
   \text{minimize} \quad H_{\textit{Ising}}(s) = \sum_{i}h_is_i + \sum_{i < j}J_{ij}s_is_j 
\end{equation}

\noindent Here, $h$ is an $n$-dimensional real-valued vector, $J$ is a real-valued $n \times n$-dimensional upper triangular matrix, and $s_i \in \{-1, 1\}$ are spin variables. Ising spin glass problems can equivalently be expressed as Quadratic Unconstrained Binary Optimization (QUBO) problems, which are of the following kind:

\begin{equation}
    \text{minimize} \quad H(x) = x^T\mathcal{Q}x = 
\sum_{i}^n\mathcal{Q}_{ii}x_i + \sum_{i <j}\mathcal{Q}_{ij}x_ix_j
\end{equation}

\noindent Here, $Q$  is a real-valued $n \times n$-dimensional upper triangular matrix and $x_i \in \{0, 1\}$ are Boolean variables. By using transformation $x_i = \frac{1}{2}(s_i +1)$, QUBO problems can be transformed to Ising spin glass problems (and vice versa). As QUBO and Ising are isomorphic and D-Wave's quantum annealers accept both these problem types as an input, they can be used interchangeably.\\
Recent research efforts thus comprise the development of techniques to express NP-hard problems as an Ising spin glass problem \cite{lucas2014ising, ikeda2019application, papalitsas2019qubo, date2021qubo} to be able to employ quantum technologies as a new practical form of optimization technique.\\\\
For this work we will focus on satisfiability problems (\cite{arora2009computational}). The satisfiability problem ({\onlysat}) of propositional logic is informally defined as follows: Given a Boolean formula, is there any assignment of the involved variables so that the formula can be reduced to ``true''? \cite{nusslein2023solving}. They were the first problems for which NP-completeness has been shown \cite{cook1971complexity}, and that turned out to be useful to prove NP-completeness of other NP problems \cite{arora2009computational}. Besides their usefulness in proofing the NP-completeness of other problems, satisfiability problems are examples of constraint satisfaction problems that are ubiquitous in domains such as artificial intelligence \cite{arora2009computational},  (software) product lines, the tracing of software dependencies, or formal methods \cite{gabor2019assessing}. Despite decades of research efforts and the invention of various different methods to solve satisfiability problems, to this day no algorithm is known that can solve arbitrary \onlytsat instances in worst-case polynomial time. Thus, \onlytsat problems have been subject to research in the context of quantum annealing as well \cite{choi2010adiabatic, choi2008minor,  chancellor2016direct, gabor2019assessing, nusslein2023solving, kruger2020quantum, sax2020approximate}. This research led to several different \onlytsatt-to-QUBO transformations (i.e. \cite{choi2010adiabatic, chancellor2016direct, nusslein2023solving})  that can all be used to solve \onlytsat problems on quantum annealers. These \onlytsatt-to-QUBO transformations differ in various properties like the dimension of the associated QUBO matrix (which is correlated with the number of needed qubits on quantum hardware), the density of the QUBO matrix (i.e. how many non-zero values are there outside of the main diagonal), the signs, positions and magnitudes of the values within the QUBO matrices. Although there are numerous transformations from \onlytsat to QUBO (or Ising) with different properties, they have not been compared with regards to the influence they have on the solution quality returned by a quantum annealer.\\\\
Hence, in this work we will study the influence that a given mapping from \onlytsat to QUBO or Ising has on the solution quality of quantum annealing. To do so, we will use four different QUBO mappings to solve 1000 hard \onlytsat instances on the D-Wave Advantage\_system4.1 and compare its returned answers. We will show that a specific \onlytsat-to-QUBO mapping can have a significant impact on the solution quality and we will provide evidence for one of the reasons that impacts the solution quality.


\section{Foundations}
In this section, we formally introduce the necessary foundations for solving \onlytsat problems on quantum annealers.




\subsection{Satisfiability Problems}
In this section we formally introduce satisfiability problems.
\begin{dfn}[Boolean formula]
A \emph{Boolean formula} over the variables $u_1,..., u_n$ consists of the variables and the logical operators \onlyand ($\wedge$), \onlyor ($\vee$), \onlynot ($\neg$). If $\varphi$ is a  Boolean formula over the variables $u_1,..., u_n$ and $z \in \{0, 1\}^n$, then $\varphi(z)$ denotes the value of $\varphi$ when each $u_i$ is assigned the value $z_i$ (where we identifiy 1 with TRUE and 0 with FALSE). A Boolean formula $\varphi$ is \emph{satisfiable} if there exists some assignment z such that $\varphi(z)$ is TRUE. Otherwise, we say that $\varphi$ is unsatisfiable. \cite{arora2009computational} 
\end{dfn}
\noindent An example for a Boolean formula over the variables $x_1, x_2, x_3$ is $\varphi_1(x_1, x_2, x_3) =  (x_1 \vee x_2 \vee x_3) \wedge (\neg x_1 \vee x_2 \vee \neg x_3)$. This formula is satisfiable as, for example, $\varphi_1(1,1,1) = $ TRUE. 
\begin{dfn}[Conjunctive Normal Form \cite{arora2009computational}] A Boolean formula over variables $u_1,...,u_n$ is in \emph{Conjunctive Normal Form (CNF form)} if it is an \emph{AND} of \emph{OR}s of variables or their negations:
$$\bigwedge_i \big(\bigvee_{j} v_{i_j}\big)$$
Each $v_{i_j}$ is either a variable $u_k$ or its negation $\neg u_k$. The $v_{i_j}$ are called the \emph{literals} of the formula and the terms $(\vee_j v_{i_j})$ are called its \emph{clauses}. A \emph{$k$CNF} ist a CNF formula, in which all clauses contain at most $k$ literals. \cite{arora2009computational}
\end{dfn}
\noindent An important special case in mathematical logic are \onlytsat problems. 
\begin{dfn}[3SAT]
A \onlytsat instance is a Boolean formula in 3CNF form, which consists of \emph{n} variables and \emph{m} clauses. The problem of deciding whether a given \onlytsat instance is satisfiable is called \onlytsatt.
\end{dfn}
\noindent It is well known (\cite{gent1994sat}) that for \onlytsat instances, the ratio of the number of clauses $m$ to the number of variables $n$ ($m/n$ ratio) is a good predictor for the hardness of the instance. For random \onlytsat instances, in which the literals of each clause are drawn uniformly from a pool of $n$ variables, potentially hard instances possess an $m/n$ ratio of approximately 4.24 \cite{gent1994sat}.


\subsection{Minor Embedding}\label{sec:minor_emb}
In all current quantum annealers produced by the company D-Wave Systems, the qubits are arranged in a specific, not fully interconnected topology. In the D-Wave \textit{Advantage\_system4.1}, which we  use in this work, this topology is called the \emph{Pegasus graph}\footnote{see further: \url{https://docs.ocean.dwavesys.com/en/stable/concepts/topology.html\#topology-sdk-pegasus}} (PG). A QUBO instance, which is an $n \times n$-dimensional matrix $Q$, can also be interpreted as a graph with $n$ vertices. The values on the main diagonal of the matrix Q denote the vertex weight, while all other elements $Q_{ij}$ with non-zero values indicate that there is an edge between vertices $i$ and $j$ and that this edge is weighted with $Q_{ij}$. With these interpretations we now have two graphs --- the QUBO graph and the topology of the quantum annealer. To solve QUBO instances on a quantum annealer, one has to find a mapping from the QUBO graph to the PG. This mapping is called a \emph{minor embedding}. We will refer to it as \emph{embedding} in the remainder of this work.\\
To create the embedding one has to solve an instance of a subgraph isomorphism problem, which is a NP-complete problem \cite{cook1971complexity}. Hence,  for practical purposes, D-Wave creates embeddings via a heuristic algorithm \cite{cai2014practical}. \\\\\ \textbf{Parameter scaling.} When embedding QUBO instances onto the hardware graph of the quantum annealer, the linear and quadratic values (i.e., the field strengths) must be scaled to a specific range. For the D-Wave \textit{Advantage\_system4.1} these ranges are: $h_{\textit{range}} = [-4,4]$ for the linear coefficients and $J_{\textit{range}} = [-1, 1]$ for quadratic coefficients. All QUBOs will thus be scaled to fit these allowed ranges. The scaling works according to the following formula from the official D-Wave documentation\footnote{\url{https://docs.dwavesys.com/docs/latest/c\_solver\_parameters.html}}: 

\begin{equation}\label{eq:dwave_scaling}
    \begin{aligned}
   h_{max} & = \big\{max \left( \frac{max(h)}{max(h_{\textit{range}})}, 0 \right),max \left( \frac{min(h)}{min(h_{\textit{range}})}, 0 \right) \big\} \\ 
   J_{max} &= \big\{max \left( \frac{max(J)}{max(J_{\textit{range}})}, 0 \right),max \left( \frac{min(J)}{min(J_{\textit{range}})}, 0 \right) \big\} \\
   Scale\_Factor &= max(h_{max}, J_{max})
    \end{aligned}
\end{equation}

\noindent To receive the scaled version of a given QUBO instance, all values of the QUBO instance will be divided by the \emph{Scale\_Factor} we just calculated.


\subsection{Maximum-Weight Independent Set}\label{found:mwis}
The Maximum-Weight Independent Set (MWIS) problem will later be used to solve \onlytsat on a quantum annealer. Hence, we review the definition of this problem here. For this whole section, let $G = (V,E)$ be a graph $G$ with vertex set \emph{V} and edge set \emph{E}. 

\begin{dfn}[Independent Set Problem]
Let $k$ be a natural number. A subset $V' \subset V$ of vertices of graph $G$ is called an independent set of $G$, if the subgraph induced by $V'$ contains no edges, i.e., if $\;  \forall v', u' \in V': (v',u') \notin E$. \\
The problem to determine whether a given graph has an independent set of cardinality $k$ is called the independent set problem. \cite{goldreich2008computational}
\end{dfn}

\noindent We can now formulate the optimization version of MWIS:
\begin{dfn}[Maximum-Weight Independent Set]
Given graph \emph{G}, where additionally each vertex $i \in V$ is weighted by a positive rational number $c_i$. The MWIS problem is to find a subset $S \subset V$ such that S is independent and the total weight of $S$ ($=\sum_{i\in S}c_i)$ is maximized. We denote the optimal set by mis(G). \cite{choi2010adiabatic}
\end{dfn}
\noindent To solve MWIS on a quantum annealer all that is left is to formulate the MWIS problem as a QUBO problem. This can be done as follows \cite{choi2008minor}:\\

\noindent If $Q_{ij} \geq \min\{c_i, c_j\}$ for all $(i,j) \in E$, then the maximum value of
$$H(x_1, ..., x_n) = \sum_{i \in V} c_ix_i - \sum_{(i,j) \in E} Q_{ij}x_ix_j$$

\noindent is the total weight of the WMIS. In particular, $mis(G) = \{i \in V: x_i^* =1\}$, where $(x_1^*, ..., x_n^*) = \argmax_{(x_1, ..., x_n) \in \{0,1 \}^n} \;\; H(x_1, ..., x_n)$. \\\\
We note that $c_i$ and $Q_{ij}$ are free parameters and can be chosen arbitrarily, as long as $Q_{ij} \geq \min\{c_i, c_j\}$ is satisfied for all $i, j \in V $.


\section{Related Work}\label{sec:related_work}
In this section, we are going to review all methods of mapping \onlytsat instances to QUBO problems that will be used in this paper. 


\subsection{Choi's Transformation}\label{sec:choi}
Choi's method \cite{choi2010adiabatic} of transforming \onlytsat instances to QUBO problems is based on a well known polynomial-time reduction from \onlytsat to MIS \cite{dasgupta2008algorithms}, which we will briefly explain here.\\
Given a \onlytsat instance $\varphi(x_1, ..., x_n) = C_1 \wedge ... \wedge C_m$ with $n$ variables and $m$ clauses and an empty graph $G_{SAT} = (V_{G_{SAT}}, E_{G_{SAT}})$:

\begin{itemize}
 \item For each clause $C_i = y_{i_1} \vee y_{i_2} \vee y_{i_3}$, we add a new 3-node clique to $G_{SAT}$. The three nodes of this 3-node clique are labeled $y_{i_1}, y_{i_2}$, and $y_{i_3}$, where each $y_{i_j} \in \{x_1, ..., x_n\} \cup \{\neg x_1, ..., \neg x_n\} $.
 \item If the labels between two vertices of different 3-node cliques are in conflict, we connect them with an edge. More precisely, for clause indices $i \neq j$ and $s,t \in \{1,2,3\}$, we add edge $(i_s,j_t)$ to graph $G_{SAT}$ iff $y_{i_s} = \neg y_{j_t}$.
\end{itemize}
Given this MWIS problem, we can create the corresponding QUBO via the transformation explained in Sec. \ref{found:mwis}. For clarity we note again that, as explained in Sec. \ref{found:mwis}, this QUBO formulation has two free parameters: the vertex weight $c_i$ and the edge weight $Q_{ij}$ for all $i,j \in V_{G_{SAT}}$.\\ As a consequence of the reduction to MWIS, the QUBO corresponding to a \onlytsat formula with $m$ clauses has dimensions $3m \times 3m$.


\subsection{Chancellor's Transformation}\label{sec:chancellor_to_qubo}
In this section, we explain Chancellor's  \onlytsat to QUBO mapping, as described in \cite{chancellor2016direct}.\\
Let $C = (x_1 \vee x_2 \vee x_3)$ be a clause of a \onlytsat instance over Boolean variables $x_1, x_2, x_3$. If we set 1$\coloneqq$True and  0$\coloneqq$False, then the following holds:
\begin{equation}\label{eq:chancellor_1}
    (x_1 \vee x_2 \vee x_3) = x_1 + x_2 + x_3 -x_1x_2 -x_1x_3 - x_2x_3 + x_1x_2x_3
\end{equation}

By using the transformation $x_i = \frac{1}{2}(s_i +1)$, where $s_i \in \{-1, 1\}$, one can move from Boolean variables to spin variables. The goal is now to formulate the right hand side of Eq. \ref{eq:chancellor_1} as an Ising problem in which all assignments of truth values to the Boolean variables that satisfy the given clause have the same optimal energy and the one assignment that does not satisfy the clause has a higher energy. The only problem with the right hand side of Eq. \ref{eq:chancellor_1} is that QUBO (and Ising) problems only allow for quadratic interactions. However, term $x_1x_2x_3$ is cubic. So all that is left to do is to find a way to express the cubic term $x_1x_2x_3$ as an Ising (or QUBO) problem and add the remaining linear and quadratic coefficients of the right hand side of Eq. \ref{eq:chancellor_1} to the Ising (or QUBO) representation of the cubic term. Chancellor describes two different methods to express the cubic term $x_1x_2x_3$ as an Ising (QUBO) problem. In our paper, we will use the first method --- a three bit parity check.\\
As we want to minimize the energy of the Ising problem, we reformulate Eq. \ref{eq:chancellor_1}, which is currently a maximization problem, into a minimization problem:
\begin{equation}\label{eq:chancellor_2}
   -(x_1 \vee x_2 \vee x_3) = -x_1 - x_2 - x_3  + x_1x_2  + x_1x_3 + x_2x_3 - x_1x_2x_3
\end{equation}
Now, each satisfying assignment has a value of $-1$, while $x_1 = x_2 = x_3 = 0 $ as the only not satisfying assignment has a value of $0$. Next, we transform the Boolean variables $x_1, x_2, x_3$ of Eq. \ref{eq:chancellor_2} to spin variables $s_1, s_2, s_3$ via transformation $x_i = \frac{1}{2}(s_i +1)$.
To model the cubic term, we need an additional qubit, also called \emph{ancilla} qubit, which we will denote $s_a$. We can now model the cubic term:

\begin{equation} \label{eq:chancellor_clause}
    \begin{aligned}
I_{\textit{cubic}}(s_{1}, s_{2}, s_{3}) = J \sum_{i=1}^3\sum_{j=1}^{i-1}c(i)c(j)s_{i}s_{j} + h \sum_{i=1}^3c(i)s_{i} + \\ J_a \sum_{i=1}^3c(i)s_{i}s_{a} + h_{a}s_{a}
    \end{aligned}
\end{equation}
In this equation, $c(i)$ is the sign of the corresponding Boolean variable $x_i$ in the clause we are trying to implement. As all the variables in the clause of Eq. \ref{eq:chancellor_2} are not negated, $c(i) = 1$.
\noindent To implement a 3 bit parity check, $h, J, J_a, h_a$ need to satisfy the following constraints: 
\begin{itemize}
    \item $h = +1$ if the sign of the cubic term in Eq. \ref{eq:chancellor_2} is $+1$, else $h = -1$.
    \item $h_a = 2h$
    \item $J_a = 2J > |h|$, which means the magnitude of $J$ can be chosen freely, as long as $2J > |h|$ holds.
\end{itemize}
To implement the clause, we only need to add the linear and quadratic coefficients of the variables on the right hand side of Eq. \ref{eq:chancellor_2} onto the formerly created Ising representation of the cubic term and we have successfully implemented a single clause as an Ising minimization problem.\\\\To implement a complete \onlytsat formula as an Ising minimization problem, one repeats the process we just described for each clause (i.e., for each clause a new ancilla qubit will be added to the formulation) and then superimposes the construction of the individual clauses on a common set of logical spin variables. That is, suppose a given \onlytsat formula consists of variables $x_1, ..., x_n$ and $m$ clauses $C_1, ... , C_m$. The corresponding Ising minimization problems will consist of spin variables $s_1, ..., s_n$, which directly correspond to Boolean variables $x_1, ..., x_n$ and $m$ additional ancilla qubits $s_{a_1}, ..., s_{a_m}$, which are needed to model the cubic term of each clause. The superimposing works as follows:
\begin{itemize}
    \item For each $s_i$: Let $K$ be the set of all clauses in which $x_i$ or $-x_i$ appears as a variable. Then, $h_i = \sum_{l=1}^{|K|}h_lc(i)_l$. Remember, the sign of $h$ might be different for different clauses, hence $h_l$ denotes the value of $h$ in the $l$-th clause of $K$. The value of $c(i)_l$ is the sign of the Boolean variable $x_i$ in $l$-th clause of $K$.
    \item Let $P$ be the set of clauses in which variables $x_i$ and $x_j$ appear together, $i < j$: $J_{ij} = \sum_{l=1}^{|P|}Jc(i)_lc(j)_l$. Again, $c(i)_l$ and $c(j)_l$ represent the sign of variable $x_i$ resp. $x_j$ in the $l$-th clause of $P$.
    \item All quadratic and linear values that include an ancilla qubit will not be modified at all. 
    
\end{itemize}
Thus the whole transformation requires $n + m$ qubits, where $n$ is the number of variables of the \onlytsat instance and $m$ is the number of clauses of the \onlytsat instance.\\\\ When choosing $J = |h|$, the resulting QUBOs will have less quadratic values than QUBOs that were created by choosing $J > |h|$. These variants are thus not isomorphic. We will use both of these versions in our evaluation later in this paper.


\subsection{Nuesslein's Transformation}
In \cite{nusslein2023solving} two novel approaches for translating arbitray \onlytsat instances to QUBO problems were presented. In this work we will use the $n+m$-approach of \cite{nusslein2023solving} and refer to it as the \textbf{Nuesslein} approach. \\
In this approach, the literals in each clause will be sorted, such that the negated literals of each clause are at the end of the clause. Thus there are only four types of clauses: one with no negated literals, one with a single negated literal at the end of the clause, one with two negated literals at the end of the clause, and one where all literals are negated. Analogously to the previously presented Chancellor approach, for each clause of the \onlytsat instance, a new pattern QUBO will be constructed. After that, one superimposes the constructed pattern QUBOs over a common set of variables, exactly as described in the Chancellor approach (see Sec. \ref{sec:chancellor_to_qubo}).
The only difference between the Chancellor and the Nuesslein approach is the structure of each of the clause QUBOs (i.e., amount and values of the different quadratic and linear values). \\
The four possible patterns for clauses of a 3-SAT formula are shown in Table~1 (where $K$ is an additional ancilla qubit that gets added per clause).

\begin{table}[!htb]
   \caption{Pattern QUBOs for the four different types of clauses \cite{nusslein2023solving}}
    \begin{minipage}{.5\linewidth}
      \caption*{(a) $( a \lor b\lor c )$, $H^* = -1$}
      \vspace{-.25cm}
\centering
\begin{tabular}[h]{|r||p{0.3cm}|p{0.3cm}|p{0.3cm}|c|}
\hline
& a & b & c & K \\
\hline
\hline
a & & 2 & & -2 \\
\hline
b & & & & -2 \\
\hline
c & & & -1 & 1 \\
\hline
K & & & & 1 \\
\hline
\end{tabular}
    \end{minipage}%
    \begin{minipage}{.5\linewidth}
      \centering
        \caption*{(b) $( a \lor b \lor \lnot c )$, $H^* = 0$}
        \vspace{-.25cm}
        \begin{tabular}[h]{|r||p{0.3cm}|p{0.3cm}|p{0.3cm}|c|}
\hline
& a & b & c & K \\
\hline
\hline
a & & 2 & & -2 \\
\hline
b & & & & -2 \\
\hline
c & & & 1 & -1 \\
\hline
K & & & & 2 \\
\hline
\end{tabular}
    \end{minipage}

\vspace{.25cm}
\begin{minipage}[!h]{.5\linewidth}

      \caption*{(c) $( a \lor \lnot b \lor \lnot c )$, $H^* = 0$}
      \vspace{-.25cm}
      \centering
      \begin{tabular} [h]{|r||p{0.3cm}|p{0.3cm}|p{0.3cm}|c|}
\hline
& a & b & c & K\\
\hline
\hline
a & 2 & -2 & & -2 \\
\hline
b & & & & 2 \\
\hline
c & & & 1 & -1 \\
\hline
K & & & & \\
\hline
\end{tabular}
    \end{minipage}%
    \begin{minipage}{.5\linewidth}
      \centering
        \caption*{(d) $( \lnot a \lor \lnot b \lor \lnot c )$, $H^* = -1$}
        \vspace{-.25cm}
       \begin{tabular}[h]{|r||p{0.3cm}|p{0.3cm}|p{0.3cm}|c|}
\hline
& a & b & c & K \\
\hline
\hline
a & -1 & 1 & 1 & 1 \\
\hline
b & & -1 & 1 & 1 \\
\hline
c & & & -1 & 1 \\
\hline
 K & & & & -1 \\
\hline
\end{tabular}
    \end{minipage} 
     
\end{table}

\section{Methodology}
In this section we describe how we created the \onlytsat instances and how we parameterized the QUBO-to-\onlytsat transformations that we used.
\subsection{Pre-Evaluation: Size of the \onlytsat Instances}
To conduct our experiments, we need to create \onlytsat instances. In this work, all \onlytsat instances will be created as follows:

\begin{itemize}
    \item Determine a ratio $m$ (= \#clauses) / $n$ (= \#variables) closely around 4.24.
    \item For each of the $m$ clauses: draw three different variables from the variable pool of size $n$ at random. For each of the chosen variables, there is a 50\% chance that they will get negated in this clause.
    \item Use a \onlytsat solver to verify that the created \onlytsat instances are solvable.
\end{itemize}
To study the influence of the chosen QUBO transformation on the solution quality, we need to ensure that the quantum annealer is able to satisfy at least some instances in some experiments. Otherwise, if the answer to every instance in every experiment is \emph{no satisfying solution found}, although we created only satisfiable \onlytsat instances, the experiments would not be helpful. Hence, we conducted a pre-evaluation, in which we solved \onlytsat instances of sizes $m=180$ and gradually decreased the size of the formulas (i.e., the amount of clauses $m$), until we were able to satisfy at least 50\% of the submitted \onlytsat problems for every QUBO transformation. This is why we chose $m = 46$ and $n = 11$ as the size of our \onlytsat instances. 

\subsection{Influence of the Embedding}
To solve QUBO problems on a quantum annealer, they need to be embedded into the hardware topology of the quantum annealer (see Sec. \ref{sec:minor_emb}). As this is done by a heuristic algorithm, generated embeddings for a given problem may differ significantly in size. In the current NISQ era, where no practical error correction is available and calculations are thus heavily influenced by noise, this may have an impact on the solution quality. Thus, there is a possibility that given multiple embeddings for the same problem, the largest embedding (i.e., the embedding that uses the most physical qubits on the hardware system) may yield worse results than an embedding that uses significantly fewer qubits to solve the same problem.\\In all our experiments, we will only create a single embedding per formula and QUBO formulation. We will not try to optimize these embeddings as this is (potentially) computationally expensive. From a problem solving perspective, there is not much benefit in being able to solve a problem with a potential speed up on a quantum hardware system, when the preparation steps (potentially) need exponential time.

\subsection{Parameterization of the QUBO Transformations}
The QUBO transformations by Choi and Chancellor (see Sec. \ref{sec:choi}, \ref{sec:chancellor_to_qubo}) allow to choose parameters freely, as long as some conditions are satisfied. Thus, in this section we clarify which parameters we have used.\\
For the \textbf{Choi} transformation, we used $c_i = -1$ as the vertex weight and $Q_{ij} = 3$ as the edge weight.\\
For the Chancellor transformation, we used two different parameterization in this paper. We call the first one ChancellorJ1 and the second one ChancellorJ5.\\
For the \textbf{ChancellorJ1} approach we choose $J =  |h| = 1$ and thus $J_a = 2$.\\
For the \textbf{ChancellorJ5} approach we choose $J = 5$ and thus $J_a = 10$.

\section{Results}
In this section we describe the experiments we have conducted, present the results of these experiments, and formulate and show a hypothesis that helps to explain the results.
\subsection{Experiment 1: Comparison of QUBO Transformations}
\begin{figure}[h]
    \centering
    \includegraphics[width=0.5\textwidth]{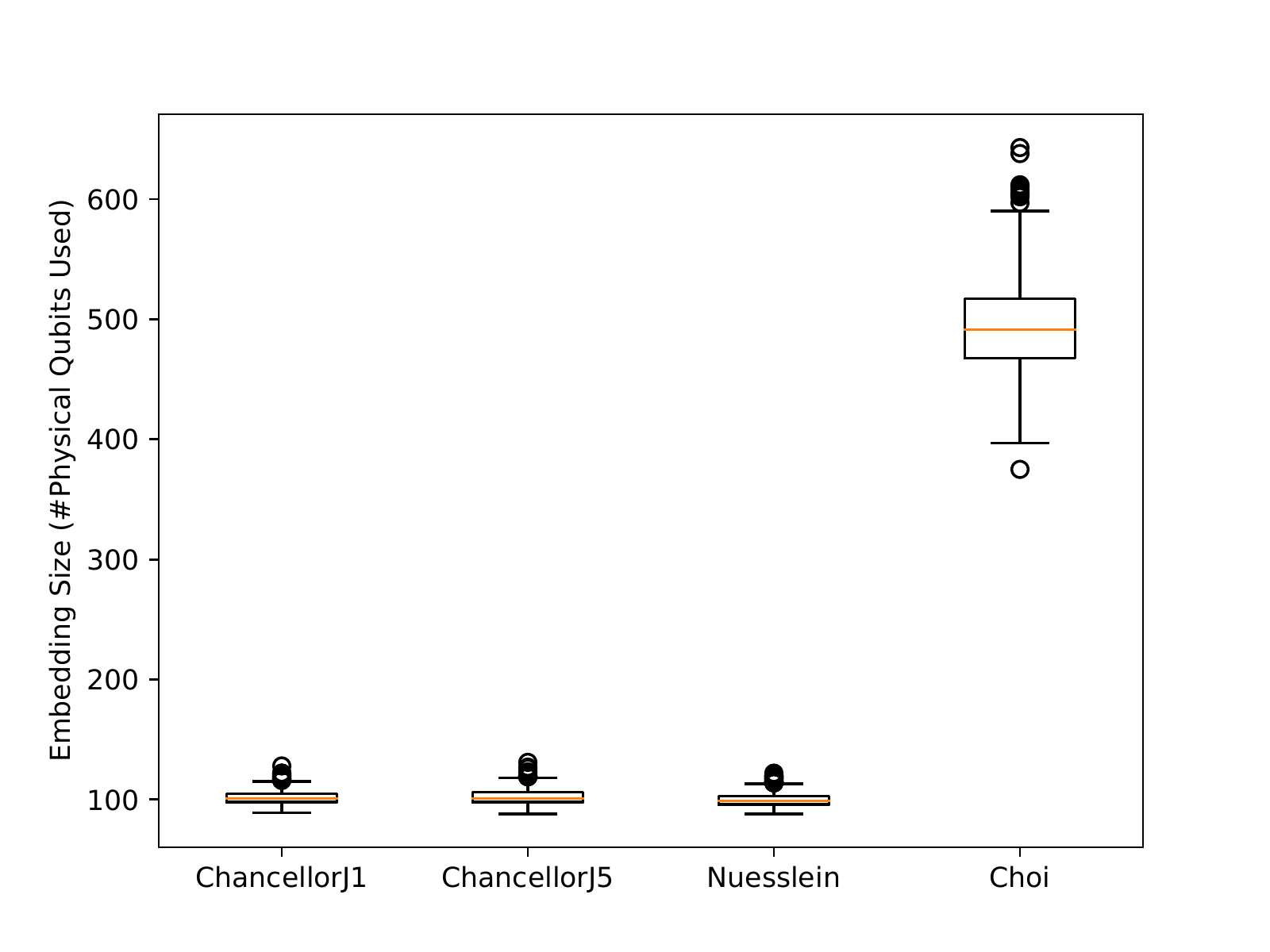}
    \caption{Embedding sizes for experiment 1. Noticeably the Choi QUBOs needed 4-6 times as many physical qubits as all the other QUBOs.}
    \label{fig:emb_exp_1}
\end{figure}

As our first experiment, we created 1000 satisfiable \onlytsat instances with $m = 46$ clauses, $n = 11$ variables (m/n $\approx$ 4.18). For each of the \onlytsat instances, we created one embedding for each QUBO resulting form the QUBO transformations Choi, ChancellorJ1, ChancellorJ5, and Nuesslein. The sizes of the generated embeddings can be seen in Figure \ref{fig:emb_exp_1}.

\noindent For each of the QUBO transformations, we solved each \onlytsat instance with the previously calculated embedding 1000 times on the D-Wave \emph{Advantage\_system4.1}. Thus, we created 4 million D-Wave samples in this experiment. \\

\subsubsection{Results}
The result of this experiment is shown in Table \ref{res_exp1_dwave}.

\begin{table}[h]
    \caption{Results of experiment 1. 1000 \onlytsat instances have been solved 1000 times on the D-Wave Advantage\_system4.1 per QUBO transformation.}
    \begin{tabular}{lcc} 
    \toprule
    QUBO  & \#solved \onlytsat instances & \#correct solutions\\ 
    \midrule 
    Choi & 695 (69.5\%) & 24634 ($\sim$2.46\%) \\
    ChancellorJ1 & 910 (91.0\%) & 87071 ($\sim$8.71\%) \\
    Nuesslein & 794 (79.4\%) & 55792 ($\sim$5.58\%)\\ 
    ChancellorJ5 & 580 (58.0\%) & 4792 ($\sim$0.48\%)\\ 
    \bottomrule
    \end{tabular}
     \label{res_exp1_dwave}
\end{table}
\noindent A surprising result is that the QUBO transformation ChancellorJ5, which uses significantly fewer qubits than the Choi transformation (see Fig. \ref{fig:emb_exp_1}), performed significantly worse than the Choi transformation. In the NISQ era, in which quantum calculations are heavily influenced by noise, one would expect that the QUBO transformation that uses the least amount of qubits would perform the best, as the probability of noise impacting the calculations is lower. The results of this experiment, however, indicate that this assumption may not be correct. \\ Another interesting result is that the performance of the three different $n+m$ transformations (ChancellorJ1, ChancellorJ5, and Nuesslein) varied significantly (ranging from $\approx 0.48 \%$ correct answers to $\approx 8.71 \%$ correct answers, see Table \ref{res_exp1_dwave}). As all these transformations use approximately the same amounts of physical qubits (as shown in Fig. \ref{fig:emb_exp_1}), the differences in the solution quality can only occur due to differences in the structure of the QUBOs. Hence, we proceed to analyze the structure of the QUBOs that were created by the $n+m$ transformations ChancellorJ1, ChancellorJ5 and Nuesslein in experiment 1.\\

\subsubsection{Analysis of the QUBO structures}
\begin{figure}[h]
    \centering
    \includegraphics[width=0.5\textwidth]{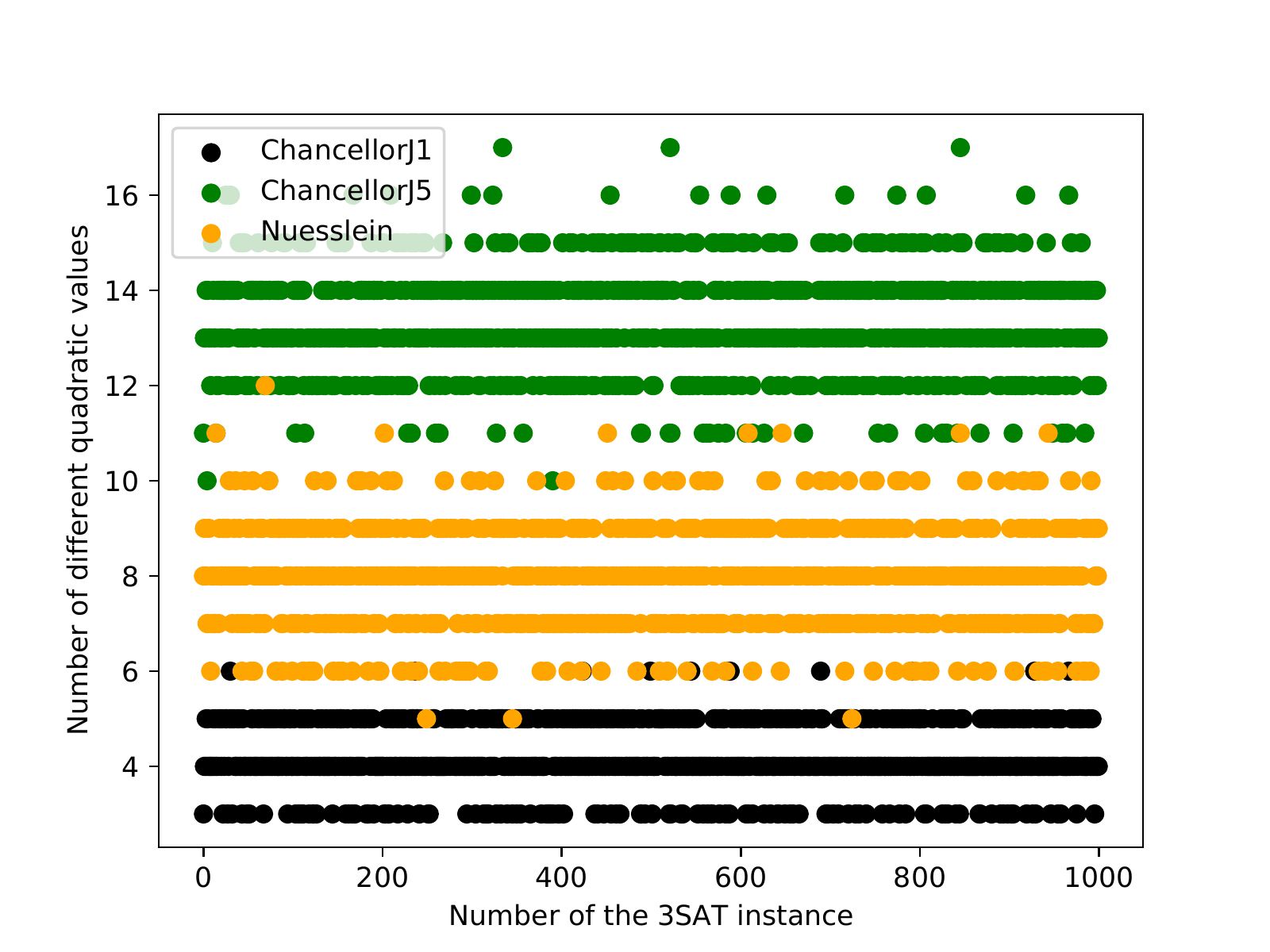}
    \caption{Number of different quadratic values for the QUBOs that were created by transformations ChancellorJ1 (black), ChancellorJ5 (green) and Nuesslein (orange) for the 1000 \onlytsat formulas of experiment1.}
    \label{fig:quadratic_couplers_1}
\end{figure}

As D-Wave scales the values of the QUBOs to a range that can be physically implemented by their quantum annealer (see Sec. \ref{sec:minor_emb}), we begin to analyze QUBO properties that are effected by scaling. These properties are the number of different non-zero quadratic and linear values as well as their ranges. \\
\noindent We begin with the analysis of the number of different quadratic values of each of the QUBOs. Therefore, for each \onlytsat formula of experiment 1 and for each of the transformations ChancellorJ1, ChancellorJ5, and Nuesslein, we count how many different quadratic values the respective QUBO contains. This result is shown in Fig. \ref{fig:quadratic_couplers_1}.

\noindent We observe that the number of different quadratic values seems to coincide with the solution quality. The QUBOs that were created by transformation ChancellorJ1 contained the least number of different quadratic values (black dots in Figure \ref{fig:quadratic_couplers_1})  and also led to the most solved \onlytsat instances and most correct solutions in experiment 1. The amount of different quadratic values and the solution quality for the transformations Nuesslein (orange dots in Figure \ref{fig:quadratic_couplers_1}) and ChancellorJ5 (green dots in Figure \ref{fig:quadratic_couplers_1}) fit this observation as well.\\ We have conducted the same analysis for the linear values of the same QUBOs. The result of this analysis however showed no meaningful correlation to the results of experiment 1. \\

\begin{figure}[h]
    \centering
    \includegraphics[width=0.5\textwidth]{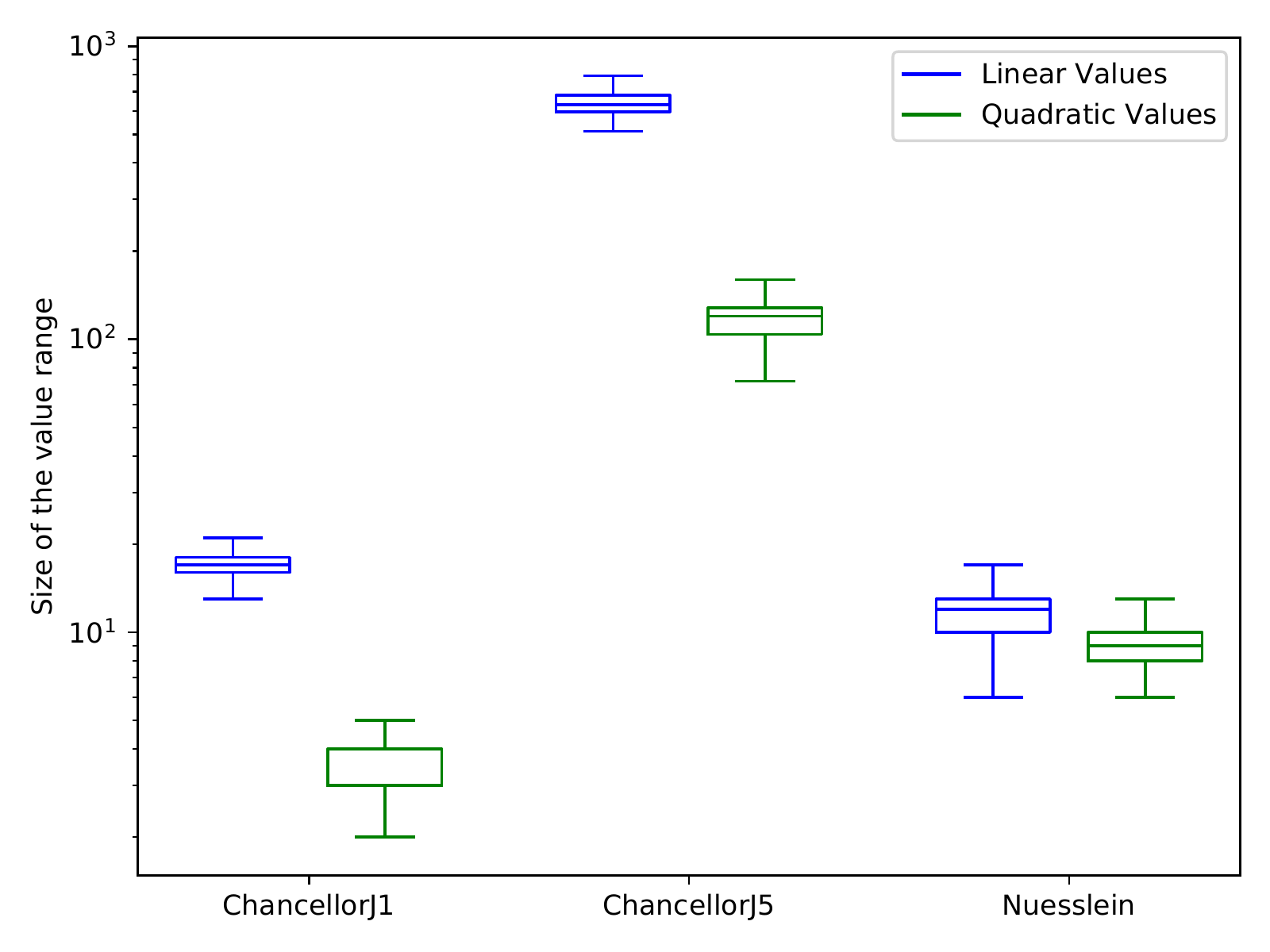}
    \caption{Distribution of the size of the values ranges (biggest quadratic (linear) value minus smallest quadratic (linear) value of a QUBO) for all QUBOs of experiment 1 (1000 per transformation).}
    \label{fig:quadratic_couplers_values_ranges}
\end{figure}

\noindent Next we analyze the size of the value ranges of the quadratic and linear values of all created QUBOs. That is, we calculate the difference between the biggest and the smallest quadratic value for each QUBO (analogously for the linear value ranges). Thus the sizes of the value ranges indicate how closely the magnitudes of all of the quadratic resp. linear values lie together. The result of the analysis is shown in Figure \ref{fig:quadratic_couplers_values_ranges}.

\noindent Note, that Fig. \ref{fig:quadratic_couplers_values_ranges} shows the distribution of the sizes of the value ranges of the quadratic resp. linear values of the analyzed QUBOs visualized by boxplots. It does not show the actual value ranges. The green boxplots in Fig. \ref{fig:quadratic_couplers_values_ranges} show the distributions of the sizes of the quadratic values and the blue boxplots show the distributions of the sizes of the linear values of the analyzed QUBOs. To make subtle differences in the presented distributions visible, the y-axis of Fig. \ref{fig:quadratic_couplers_values_ranges} is in log scale.

\noindent We observe that the sizes of the value ranges of the quadratic values  of the ChancellorJ1 QUBO  are the smallest, the value ranges of the quadratic values of the Nuesslein QUBOs are the second smallest and the value ranges of the ChancellorJ5 QUBOs are the largest. Thus for this analysis we can see a correlation between the size of the values ranges of the quadratic values of the analyzed QUBOs and the results of experiments 1. As we have seen in Fig. \ref{fig:quadratic_couplers_values_ranges}, the ChancellorJ1 QUBOs possess the smallest size of the value ranges of their quadratic values. The ChancellorJ1 QUBOs also performed the best in experiment 1. The QUBOs created by the Nuesslein and ChancellorJ5 QUBOs fit this observation perfectly.\\
With regards to the size of the value ranges of the linear values (blue boxplots in Fig. \ref{fig:quadratic_couplers_values_ranges}) of the created QUBOs, we do not see the same correlation. We note, that ChancellorJ5 QUBOs possessed the biggest size of the value range of the linear value and also performed the worst in experiment 1. However, Nuesslein QUBOs possessed the smallest size of the value range of the linear values but did not perform best (but second best) in experiment 1 and ChancellorJ1 QUBOs possessed the second smallest size of the value range of the linear values, but did not perform second best (but best) in experiment 1.

\subsubsection{Possible explanation for the  difference in solution quality}

After comparing differently structured QUBOs of the same size, as is the case with the transformations ChancellorJ1, ChancellorJ5, and Nuesslein, we found an interesting correlation between the number of different quadratic values alongside their sizes of the value range and their performance in quantum annealing in experiment 1. We suggest that the more different quadratic values are present and the larger the differences between those values are, the worse the results will get.\\ In the context of applied quantum annealing, all linear and quadratic values of a QUBO instance will get scaled down to a certain range. For demonstration purposes, we pick concrete values of a ChancellorJ5 QUBO of experiment 1. This QUBO had a scaling factor of 172, which means, that all QUBO values will be divided by 172 when it is solved on the D-Wave quantum annealer. The same QUBO also had quadratic values that were very close to each other (which is very typical for all the \onlytsat to QUBO transformations we investigated in this paper). One quadratic value was 24, another one was 16. Now when scaling these values, their difference, $24/172 - 16/172 = 8/172 = 0.0465$ gets very small. The bigger the size of the value range (i.e. the larger the difference between the largest and the smallest quadratic resp. linear value) the bigger the scaling factor gets, and the more floating point precision is needed to tell subtle differences in quadratic values apart (see the formerly mentioned example) or to tell small quadratic values (before scaling) apart from zero (after scaling).
 Apart from the fact that it is unclear up to which decimal precision these values can be accurately transformed into physical field strengths on a quantum annealer, these differences may be too subtle to be held apart by the quantum annealer for the entirety of the annealing process.\\

\noindent To verify this hypothesis, we conduct two more experiments, which we will now explain.

\subsection{Experiment 2: Scaled QUBOs}
In this experiment, we use the same \onlytsat instances that we used in experiment 1 again. We also re-use the embeddings of experiment 1. However, we only use transformation ChancellorJ1. We use the QUBOs we created in experiment 1 by the ChancellorJ1 transformation, but multiply each value of each of the 1000 QUBOs by a factor of 1500. This should not have any impact on the received solution quality when solving these scaled QUBOs, as it does not change the structure of the QUBO (i.e., it does not change the number of different linear and quadratic values) and according to D-Wave's scaling mechanism (see Sec. \ref{sec:minor_emb}), D-Wave will scale these new QUBO values back down to the exact same values of the QUBOs that were used in experiment 1. However, we would like to verify this empirically, as it also improves the statistical certainty for the upcoming experiment 3. The results of this experiment are shown in Table \ref{res_exp2_dwave}.

\begin{table}[h]
    \caption{Results of experiment 2. 1000 \onlytsat instances have been solved 1000 times on the D-Wave Advantage\_system4.1 with the scaled version of the ChancellorJ1 transformation}
    \begin{tabular}{lcc} 
    \toprule
    QUBO  & \#solved \onlytsat instances & \#correct solutions\\ 
    \midrule 
    S. ChancellorJ1 & 912 (91.2) & 87180 ($\sim$8.72\%) \\
    \bottomrule
    \end{tabular}
     \label{res_exp2_dwave}
\end{table}
\noindent As expected, the results did not change significantly in comparison to the results of experiment 1 (see Table \ref{res_exp1_dwave}). We conclude that as expected, a simple scaling of all the QUBO values within a QUBO does not impact the solution quality at all.

\subsection{Experiment 3: Increased number of different quadratic values and bigger value range}
In this experiment we again use the same \onlytsat instances that we used in experiment 1. However, we are now going to increase the number of different quadratic values in the QUBOs that result from the ChancellorJ1 transformation. To do so, we refer again to the details of the construction of QUBOs according to Chancellor in Section \ref{sec:chancellor_to_qubo}. The QUBOs of our modified ChancellorJ1 version are construced as follows:
\begin{itemize}
    \item[(i)] For each clause, create a clause QUBO as explained in Section \ref{sec:chancellor_to_qubo}.
    \item[(ii)] For each of the clause QUBOs, choose a value from \{1, 500, 1001\} at random and multiply all of the values of the clause QUBO by the chosen number.
    \item[(iii)] Superimpose these modified clause QUBOs as explained in Section \ref{sec:chancellor_to_qubo}.
\end{itemize}
The values 1, 500, and 1001 are chosen arbitrarily. Their only purpose is to ensure that the number of different quadratic values increases and that the size of the value range of the quadratic values will get larger (i.e., that there are some smaller and some larger quadratic values). 
We chose 1001 instead of 1000, because 1000 is a small multiple of 500 and would thus not increase the size of the value range of the quadratic values as much as desired. \\\\
We analyzed all 1000 modified ChancellorJ1 QUBOs with regards to their number of different quadratic values alongside their corresponding size of the value range. After the modification, all 1000 QUBOs had between 8 and 18 different quadratic values (with an average of approx. 13 different quadratic values), which is approximately threefold the number of different quadratic values the QUBOs had before modification (see black dots in Fig. \ref{fig:quadratic_couplers_1}). The size of the value range of the quadratic values is now between 1500 and 5504 with an average of approximately 2765. This is approximately 1000 times the size of the value range of quadratic values the QUBOs had before scaling (see green boxplot for ChancellorJ1 in Fig. \ref{fig:quadratic_couplers_values_ranges}). For completeness, we note, that the sizes of the value ranges of the linear values of the modified ChancellorJ1  QUBOs are now between 6000 and 17000 with an average of approx. 10000. This is approx. 500 - 1000 times the size it was before scaling (see blue boxplot for ChancellorJ1 in Fig. \ref{fig:quadratic_couplers_values_ranges}. \\

\noindent The formerly described construction of the modified ChancellorJ1 QUBOs does not impact the correctness of the QUBO transformation. As outlined in Section \ref{sec:chancellor_to_qubo}, all satisfying assignments of a clause will have the same energy, while only the one assignment not satisfying the clause will have a higher energy. By multiplying each value of a clause QUBO with a positive constant $k$, we only shift the energies. Assume, for example, all satisfying assignments of a clause had energy $-p$. After scaling, all satisfying assignments now still have the lowest energy possible for the clause QUBO --- but the value now is $-kp$. Hence, all assignments satisfying a \onlytsat instance still have the equal lowest energy value --- we just shifted the magnitude. This can also easily be verified by comparing the energies of all possible variable assignments for small \onlytsat instances.\\\\
\noindent We now use this modified ChancellorJ1 transformation to solve the 1000 \onlytsat instances from experiment 1 again. We again use the same embeddings that we created in experiment 1. The only thing that has changed are thus the magnitudes of the linear and quadratic values. The result of this experiment is shown in Table \ref{res_exp3_dwave}.

\begin{table}[h]
    \caption{Results of experiment 3. 1000 \onlytsat instances have been solved 1000 times on the D-Wave Advantage\_system4.1 with the modified ChancellorJ1 transformation.}
    \begin{tabular}{lcc} 
    \toprule
    QUBO  & \#solved \onlytsat instances & \#correct solutions\\ 
    \midrule 
    Mod. ChancellorJ1 & 799 (79.9) & 28739 ($\sim$2.87\%) \\
    \bottomrule
    \end{tabular}
     \label{res_exp3_dwave}
\end{table}

\noindent In comparison to the results of experiment 1 and experiment 2, we observe that significantly fewer \onlytsat instances have been solved and even fewer correct answers have been given. As we solved the same set of \onlytsat instances twice before (once in experiment 1 and once in experiment 2) and saw that the amount of solved \onlytsat instances and correct solutions are approximately the same, the results of this experiment can be attributed to the modification of the QUBO's values. This provides empirical evidence for our formerly presented hypothesis that the number of different qudaratic values alongside the size of their value range impacts the solution quality of quantum annealing.

\section{Discussion}
From an optimization perspective, the modification of the QUBO in experiment 3 introduced a bias: by increasing the rewards and penalties by the factors 500 and 1001, we incentivize the optimizer heavily to make sure the clauses with the highest rewarding linear and quadratic values are satisfied first, while less rewarded clauses will be satisfied later. For our purposes, however, this is not a problem. Our modification did not change the optima. Specifically, the optimal bit strings (i.e., the assignments satisfying the \onlytsat instance) of the modified ChancellorJ1 QUBO are exactly the same as the optimal bit strings of the unmodified ChancellorJ1 QUBO --- it is only the magnitude that is different. Although we produced this phenomenon artificially in experiment 3, QUBO transformations often come with free parameters (see Choi and Chancellor transformations in this paper) --- and as we have shown in our work (i.e. through the results of experiment 1, where the differently parameterized Chancellor transformations ChancellorJ1 and ChancellorJ5 performed significantly different), the choice of these parameters is very important.  \\Finally, we have only discussed the differences between the three $n+m$ approaches and left the Choi approach unnoticed. The sizes of the values ranges of the quadratic and linear values of the QUBOs produced by Choi transformation are one. That is, everything is rewarded equally (by the first free parameter of the transformation) and everything is penalized equally (by the second free parameter of this transformation). Thus the number of different linear and quadratic values as well as the sizes of the values ranges of the QUBOs produced by the Choi transformation are smaller than the same properties of QUBOs produced by the ChancellorJ1 and Nuesslein transformation, which both yielded better results in experiment 1 than the Choi transformation nevertheless. We thus conclude that the number of different linear and quadratic values alongside their respective ranges are only one part of the explanation with regards to the performance of a QUBO transformation. At some point other factors, like the number of physical qubits needed (and thus the susceptibility to noise) or other yet unknown factors or interactions need to be considered.
\section{Conclusion}
This paper we showed that the choice of a \onlytsat to QUBO transformation can impact the solution quality of quantum annealing significantly. We showed this empirically, by conducting a benchmark study, in which we created 6 million samples of D-Wave's quantum annealer Advantage\_system4.1. We used four different \onlytsat-to-QUBO transformations to transform 1000 hard \onlytsat instances into QUBO instances, solved these QUBO instances on the quantum annealer and compared the results. We observed, that the results between these four QUBO transformations varied up to a whole order of magnitude in the amount of correct solutions found ($\approx0.48$ to $\approx 8.71$). Furthermore we also observed, that a QUBO transformation of size $n+m$ (ChancellorJ5) produced worse results than another QUBO transformation of size $3m$ (Choi), which is significantly larger then $n+m$ in our cases. Thus the number of qubits needed by a QUBO transformation is not a good predictor of its solution quality for \onlytsat problems. We showed empirically that the number of different quadratic values of a QUBO as well as the size of their value ranges does impact the solution quality. A higher amount of different quadratic values and a larger size of the value range of the quadratic values  of a QUBO lead to worse results in quantum annealing.\\ For \onlytsat problems we want to investigate in the future, whether there are Pareto-optimal QUBO transformations for quantum annealing with regards to the size of the QUBO and the amount of different quadratic values and the size of the value range of the quadratic values. Another interesting question is, whether there are similar differences in the solution quality of quantum annealing when solving other problems with the help of multiple different QUBO transformations and if so, whether these differences can be attributed to similar reasons as the ones we described in this paper.

\begin{acks}
This paper was partially funded by the German Federal Ministry of Education and Research through the funding program ``quantum technologies --- from basic research to market'' (contract number: 13N16196).
\end{acks}

\bibliographystyle{ACM-Reference-Format}
\bibliography{sample-base}

\end{document}